# Phase Control in a Spin-Triplet SQUID

Joseph A. Glick[1], Victor Aguilar[1], Adel B. Gougam[1,2], Bethany M. Niedzielski[1†], E.C. Gingrich[3], Reza Loloee[1], W.P. Pratt, Jr.[1], and Norman O. Birge[1*]

[1]Michigan State University, East Lansing, MI 48824.
[2]Khalifa University of Science and Technology/Masdar Institute, Abu Dhabi, UAE.
[3]Northrop Grumman Corporation, Linthicum Heights, MD, 21090.
*Correspondence to birge@pa.msu.edu.
†Current address: MIT Lincoln Laboratory, 244 Wood St, Lexington, MA 02421.

**Abstract:**

It is now well established that a Josephson junction made from conventional spin-singlet superconductors containing ferromagnetic layers can carry spin-triplet supercurrent under certain conditions. The first experimental signature of that fact is the propagation of such supercurrent over long distances through strong ferromagnetic materials. Surprisingly, one of the most salient predictions of the theory has yet to be verified experimentally – namely that a Josephson junction containing three magnetic layers with coplanar magnetizations should exhibit a ground-state phase shift of either zero or π depending on the relative orientations of those magnetizations. Here we demonstrate this property using Josephson junctions containing three different types of magnetic layers, chosen so that the magnetization of one layer can be switched by 180° without disturbing the other two. Phase-sensitive detection is accomplished using a superconducting quantum interference device, or SQUID. Such a phase-controllable junction could be used as the memory element in a fully-superconducting computer.

**Main Text:**

The prediction in 2001 that spin-triplet supercurrent can be generated from conventional spin-singlet superconductors [1,2] caused considerable excitement in the field. Spin-triplet superconductors occur very rarely in nature, so the theory suggested a completely new route to the generation of spin-triplet electron pairs. The mechanism of the transformation from spin-singlet to spin-triplet pairs is now well-established conceptually [3]. Spin-singlet pairs traversing a ferromagnetic material undergo a phase shift between the up-down and down-up parts of the spin-singlet wavefunction, which generates the $m_s=0$ component of the spin-triplet state, where $m_s$ is the spin projection along the magnetization axis. If the electrons are then subjected to a rotation of the magnetization axis, $m_s=\pm 1$ triplet components are generated in the rotated basis. Before 2001, it was thought that any proximity effects or supercurrents in superconducting-ferromagnetic hybrid systems would be very short-ranged, due to the large energy and momentum shift between the majority and minority spin bands of a ferromagnetic material [4,5]. Spin-triplet pairs with projection $m_s=\pm 1$ along the magnetization axis (often called "equal-spin triplets" in the literature) overcome that problem because the two electrons propagate in the same spin band, hence they perceive the ferromagnetic material as though it were a normal metal. As a result, supercurrent carried by spin-triplet pairs can propagate long distances through ferromagnetic materials.

Experimental verification of the theory appeared initially in 2006 [6,7]; then in 2010 several groups produced overwhelming evidence for long-range supercurrents in Josephson junctions containing strong ferromagnetic materials – a telltale signature of spin-triplet electron pairs [8-11]. Our own approach to this field uses the Josephson junction design first suggested by Houzet and Buzdin in 2006 [12], in which the junction contains three ferromagnetic (F) materials with mutually perpendicular magnetizations between adjacent F layers. We showed recently that the spin-triplet supercurrent can be turned on and off by rotating the magnetization of one of those layers so that it is either parallel or perpendicular to the magnetization of the adjacent layer [13]. While the "free" magnetic layer in those samples could also be rotated by 180°, the experiment measured only the amplitude of the critical current, without garnering any information about the ground-state phase across the junction.

In this work we measure the junction phase by constructing a superconducting quantum interference device (SQUID) containing two Josephson junctions fabricated simultaneously – see Figure 1. The two junctions have different shapes – one elliptical and the other an elongated hexagon – so that the magnetization direction of the free layer in the elliptical junction can be switched without altering any of the magnetic layers in the other junction. The layer structure of the junctions is S/N/F'/N/F/N/F"/N/S, where S= superconductor, N=normal metal, and F', F, and F" are ferromagnetic materials. The two requirements for this experiment are that the magnetizations of any two adjacent layers be orthogonal to maximize generation of the $m_s = \pm 1$ triplet components, and that the magnetization of one layer – either the F' or F" layer – be free to rotate by 180° without disturbing the other two. We accomplish that by using magnetic materials with in-plane magnetization for F' and F" and out-of-plane magnetization for F. For F" we choose Ni because it is very "hard" – i.e. it requires a high field to switch its magnetization direction. For F' we choose permalloy (Py), a NiFe alloy, because it is very "soft" and switches its magnetization direction with an applied field of only a few mT. For F, we use a $[Pd/Co]_n$ multilayer (*n* repeats of a Pd/Co bilayer) with strong perpendicular magnetic anisotropy (PMA). This choice guarantees that the magnetization of F remains perpendicular to the magnetizations of F' and F" under all conditions. To reduce the possibly deleterious effect of stray magnetic fields emanating from domain walls in F, we use two back-to-back $[Pd/Co]_n$ multilayers whose magnetizations are coupled antiparallel to each other by a thin Ru spacer, to form a synthetic antiferromagnet (SAF).

The first step in the experiment is to verify that the Josephson junctions described above actually carry spin-triplet supercurrent. We verified that in a recent work [14] by comparing the supercurrent amplitude in junctions with and without the outer F' and F" layers. Junctions without those layers are expected to carry only spin-singlet supercurrent, whereas junctions with those layers should carry long-range spin-triplet supercurrent. Indeed, we found that the magnitude of the supercurrent in the latter set of samples decreased less rapidly as a function of the number *n* of bilayers in the $[Pd/Co]_n$ multilayers, confirming the long-range nature of the supercurrent in the central multilayer. In this work we focus on junctions with *n* = 2 or 3 on each side of the SAF, corresponding to a total number of [Pd/Co] bilayers of 4 or 6.

The next step is to perform the phase-sensitive experiment depicted in the bottom of Figure 1. Before starting any measurements, a large in-plane field $\mu_0 H_{set}$ = -150 mT is applied in the negative field

direction – i.e. to the left in Figure 1 – and then removed. That field initializes the magnetization directions of the Ni and Py layers in both junctions [14]. The PMA SAF is stiff enough that its magnetization hardly rotates in the initialization field, and in any case it returns to its perpendicular magnetic state after $H_{set}$ returns to zero. Then we measure the critical current $I_c$ of the SQUID as a function of the current $I_{flux}$ passing through a nearby superconducting line. (In an idealized Josephson junction, $I_c$ is defined as the largest supercurrent that can pass through the junction without causing any voltage drop to appear. Junctions with small critical current exhibit some voltage drop even when $I < I_c$ due to finite temperature and environmental noise; Figure 2(B) shows typical I-V curves along with fits to a theory that takes those into account.) The current $I_{flux}$ produces a very small out-of-plane field which induces magnetic flux through the SQUID loop. The $I_c(I_{flux})$ data exhibit oscillations with a period of about 1.5 mA, corresponding to one flux quantum $\Phi_0 = h/2e$ through the SQUID loop. We then apply a small in-plane "set field", $\mu_0 H_{set} = 0.4$ mT, in the positive direction – to the right in Figure 1. After returning the field to zero, we measure the full SQUID oscillation data $I_c(I_{flux})$ again. We repeat this sequence while increasing the magnitude of $\mu_0 H_{set}$ in steps of 0.4 mT. The results are shown for SQUID sample 2A-4 in Figure 2(A) as a 3D plot of $I_c$ vs $I_{flux}$ and $H_{set}$. The 3D plot shows that the $I_c(I_{flux})$ oscillation curve does not change for values of $\mu_0 H_{set}$ up to 2.0 mT. At $\mu_0 H_{set} = 2.4$ mT, the SQUID oscillation curve suddenly shifts by almost exactly ½$\Phi_0$, indicating that one of the junctions in the SQUID has acquired a $\pi$ phase shift relative to the initial state. According to the theory of spin-triplet Josephson junctions [2,12,15,16], that phase shift occurs when the magnetizations of the F' and F'' layers switch from being parallel to antiparallel, which is consistent with the Py F' layer in one of the junctions (probably the elliptical one) reversing its magnetization direction. We then repeat the whole procedure with negative values of $\mu_0 H_{set}$ to return the system to its initial state. As shown in Figure 2(A), the transition back to the initial state occurs at $\mu_0 H_{set} = -2.8$ mT. The flux shift of ½$\Phi_0$ is instantly apparent in the raw data without any further analysis.

To test the reproducibility and robustness of the results, we have switched device 2A-4 between its two magnetic states one thousand times, while keeping $I_{flux}$ fixed at -0.2 mA to maximize the difference between the values of $I_c$ in the two states. Figure 2(C) shows a histogram of the resulting $I_c$ values. The narrow distributions of $I_c$ values in the two magnetic states show that the behavior is highly reproducible over multiple switches. We have carried out similar measurements on eight different SQUIDs – four with *n* = 2 and four with *n* = 3. We obtained results similar to those shown in Figure 2 from seven of the eight SQUIDs; only one SQUID with *n* = 3 exhibited poor magnetic behavior with the phase appearing to move continuously rather than switching abruptly. The SQUIDs with *n*=3 have lower values of the critical current, as expected from our previous work [14]. But their phase behavior, with the exception of the one poor sample, is just as robust as that of the *n*=2 SQUIDs.

One can extract quantitative estimates of the individual junction critical currents and the SQUID phase shift by fitting the $I_c(I_{flux})$ data to standard SQUID theory, as shown in Figure 3. (Fitting details are provided in the Supplementary Materials). We summarize the two most important results of that procedure here. First, the phase shift between the two magnetic states shown in Fig. 3 differs from $\pi$ typically by a few percent. This is undoubtedly due to the change in magnetic flux induced in the SQUID by the magnetization of the Py free layer [17]. If desired, that contribution could be removed by designing a SQUID with higher symmetry, such that the magnetizations of the ferromagnetic layers in the junctions do not inject any flux into the SQUID. Second, the theory of spin-triplet Josephson junctions predicts that the amplitude of the critical current should be the same in the two magnetic states if the free layer switches its magnetization direction by exactly 180° [2,12,15,16]. Our samples exhibited that critical current symmetry only once, in SQUID 2A-1, and even that sample violated the symmetry during subsequent measurement runs. To explain that discrepancy, we initially hypothesized that our junctions might be carrying a small amount of spin-singlet supercurrent in addition to the spin-triplet supercurrent, so that the two interfere constructively in one state and destructively in the other state. But that hypothesis fails for the samples with *n*=3, where we know that the amplitude of the spin-singlet supercurrent is negligible compared to the spin-triplet supercurrent [14]. Another possible explanation is that the Fraunhofer pattern of the elliptical switching junction is shifted more in the initial state where the Ni and Py magnetizations are parallel, compared to the second state where those magnetizations are antiparallel. As a result, the measured critical current of that junction is less than the maximum value one would find at the peak of its Fraunhofer pattern [14].

A Josephson junction whose ground-state phase difference can be controllably switched between 0 and $\pi$ has potential uses in high-speed superconducting single-flux quantum circuits or in quantum computing circuits [18-23]. Our primary interest is to make a superconducting memory [17,24]. For the latter application, we envisage a SQUID loop containing two conventional SIS Josephson junctions (where I=insulator) and one ferromagnetic junction that acts as a passive phase shifter. The shift of the $I_c$(flux) curve of the SQUID is easily detected by applying appropriate current bias and flux to the memory cell. The advantage of such a design is that only the SIS junctions switch into the voltage state during readout of the memory [24]. By switching the SIS junctions, a much larger signal is generated since ferromagnetic junctions typically have an $I_c$R value of only a few $\mu$V or less.

From an applications perspective, there are easier ways to make a phase-controllable Josephson junction: we showed recently that a "spin-valve" Josephson junction containing only two magnetic layers of appropriate thicknesses could also exhibit controllable 0-$\pi$ switching [17]. In those devices, the physical mechanism of the 0-$\pi$ phase shift is different; it relies on the accurate tuning of the thicknesses of the two magnetic layers so that the total phase shift acquired by an electron pair traversing the sample is closer to an even or odd multiple of $\pi$ when the two magnetizations are parallel or antiparallel. In the spin-triplet devices presented here, the 0-$\pi$ switching is caused by spin rotations rather than phase accumulation, so the behavior is less sensitive to the exact thicknesses of the F' and F'' layers. We believe that fact partially explains the high degree of consistency we observed in the seven samples measured (see Supplementary Materials). Another possible advantage of the design presented here is that the central PMA SAF shields the free Py layer from stray magnetic fields emanating from the Ni fixed layer. A disadvantage of our devices is that their critical currents are very small. It should be possible to enhance those by optimizing the materials in the stack.

**References and Notes:**

1. F. S. Bergeret, A. F. Volkov, and K. B. Efetov, Phys. Rev. B **64**, 134506 (2001).
2. A.F. Volkov, F. S. Bergeret, and K. B. Efetov, Phys. Rev. Lett. **90**, 117006 (2003).
3. M. Eschrig, Physics Today, 64(1), 43 (2011) http://dx.doi.org/10.1063/1.3541944.
4. A.I. Buzdin, L. N. Bulaevskii, and S. V. Panyukov, Pis'ma Zh. Eksp. Teor. Fiz. **35**, 147 (1982), [JETP Lett. **35**, 178 (1982)].
5. E. A. Demler, G. B. Arnold, and M. R. Beasley, Phys. Rev. B **55**, 15174 (1997).
6. R.S. Keizer, S.T.B. Goennenwein, T.M. Klapwijk, G. Miao, G. Xiao, and A. Gupta, Nature (London) **439**, 825 (2006).
7. H. Sosnin, H. Cho, V.T. Petrashov, and A.F. Volkov, Phys. Rev. Lett. **96**, 157002 (2006).
8. T.S. Khaire, M.A. Khasawneh, W.P. Pratt Jr. and N.O. Birge, Phys. Rev. Lett. **104**, 137002 (2010).
9. J.W.A. Robinson, J.D.S. Witt and M.G. Blamire, Science **329**, 59 (2010).
10. D. Sprungmann, K.Westerholt, H. Zabel, M.Weides and H. Kohlstedt, Phys. Rev. B **82**, 060505 (2010).
11. M. S. Anwar, F. Czeschka, M. Hesselberth, M. Porcu, and J. Aarts, Phys. Rev. B **82**, 100501 (2010).
12. M. Houzet and A.I. Buzdin, Phys. Rev. B **76**, 060504(R) (2007).
13. W. Martinez, W.P. Pratt, Jr., and N.O. Birge, Phys. Rev. Lett. **116**, 077001 (2016).
14. J.A. Glick, S. Edwards, D. Korucu, V. Aguilar, B.M. Niedzielski, R. Loloee, W.P. Pratt, Jr., and N.O. Birge, Phys. Rev. B **96**, 224515 (2017).
15. A.F. Volkov and K. B. Efetov, Phys. Rev. B **81**, 144522 (2010).
16. L. Trifunovic and Z. Radovic, Phys. Rev. B **82**, 020505 (2010).
17. E.C. Gingrich, B.M. Niedzielski, J.A. Glick, Y. Wang, D.L. Miller, R. Loloee, W. P. Pratt, Jr., and N.O. Birge, Nat. Phys. **12**, 564 (2016); doi:10.1038/nphys3681.
18. E. Terzioglu, and M.R. Beasley, IEEE Trans. Appl. Supercond. **8**, 48 (1998).
19. L.B. Ioffe, V.B. Geshkenbein, M.V. Feigel'man, A.L. Fauchere, and G. Blatter, Nature **398**, 679 (1999).
20. A.V. Ustinov, V.K. Kaplunenko, J. Appl. Phys. **94**, 5405 (2003).
21. T. Yamashita, K. Tanikawa, S. Takahashi, and S. Maekawa, Phys. Rev. Lett. **95**, 097001 (2005).
22. M.I. Khabipov *et al.*, Sci. Technol. **23**, 045032 (2010).
23. A.K. Feofanov *et al.*, Nat. Phys. **6**, 593-597 (2010).
24. I.M. Dayton *et al.*, IEEE Magnetics Letters (2018), DOI: 10.1109/LMAG.2018.2801820.
25. B. Chesca, R. Kleiner, and D. Koelle, SQUID Theory. Ch. 2 of The SQUID Handbook. Vol. 1, Clarke J & Braginski AI (Eds), WILEY-VCH Verlag GmbH & Co. KGaA, Weinheim (2004).
26. D. L. Edmunds, W. P. Pratt, and J. A. Rowlands, Rev. Sci. Instrum. **51**, 1516 (1980).
27. A. Barone and G. Paterno, Physics and applications of the Josephson effect (Wiley, 1982).
28. Yu. M. Ivanchenko and L. A. Zil'berman, Zh. Eksp. Teor. Fiz. **55**, 2395 (1968), [Sov. Phys. JETP **28**, 1272 (1969)].
29. V. Ambegaokar and B. I. Halperin, Phys. Rev. Lett. **22**, 1364 (1969).
30. J.A. Glick, PhD thesis, Michigan State University (2017).

**Acknowledgments:** We thank A. Herr, D. Miller, O. Namaan, N. Rizzo, and M. Schneider for helpful discussions, J. Willard for performing FastHenry simulations, and B. Bi for help with fabrication using the Keck Microfabrication Facility. This research is supported by the Office of the Director of National Intelligence (ODNI), Intelligence Advanced Research Projects Activity (IARPA), via U.S. Army Research Office Contract No.W911NF-14-C-0115. The views and conclusions contained herein are those of the authors and should not be interpreted as necessarily representing the official policies or endorsements, either expressed or implied, of the ODNI, IARPA, or the U.S. Government.

**Fig. 1.** Spin-triplet Josephson junction structure and SQUID loop design. (Top) Schematic cross-section of the central layers in our Josephson junctions (not to scale). The central F layer is composed of two [Pd (0.9 nm)/Co (0.3 nm)]$_n$ multilayers with perpendicular magnetic anisotropy (PMA), separated by a Ru (0.95 nm) spacer to form a synthetic anti-ferromagnet (SAF). The outer F′ and F″ layers have in-plane magnetization; we used Permalloy (Py) for F″ and Ni for F′. One junction has an elliptical cross-section (aspect ratio 2.0) to make its F′ layer switch at a low field, while the other is an elongated hexagon (aspect ratio 3.0); both have area 0.5 μm$^2$. (Bottom) The two junctions are arranged into a SQUID loop. An external field $H_{set}$ is used to control the magnetization directions of the F′ and F″ layers inside the junctions; all measurements are performed with $H_{set}$=0. The current $I_{flux}$ passing through a nearby superconducting line creates an out-of-plane field $H_{flux}$, which couples magnetic flux into the SQUID loop. The Py magnetizations are shown as black arrows labeled $M_{Py,1}$ and $M_{Py,2}$.

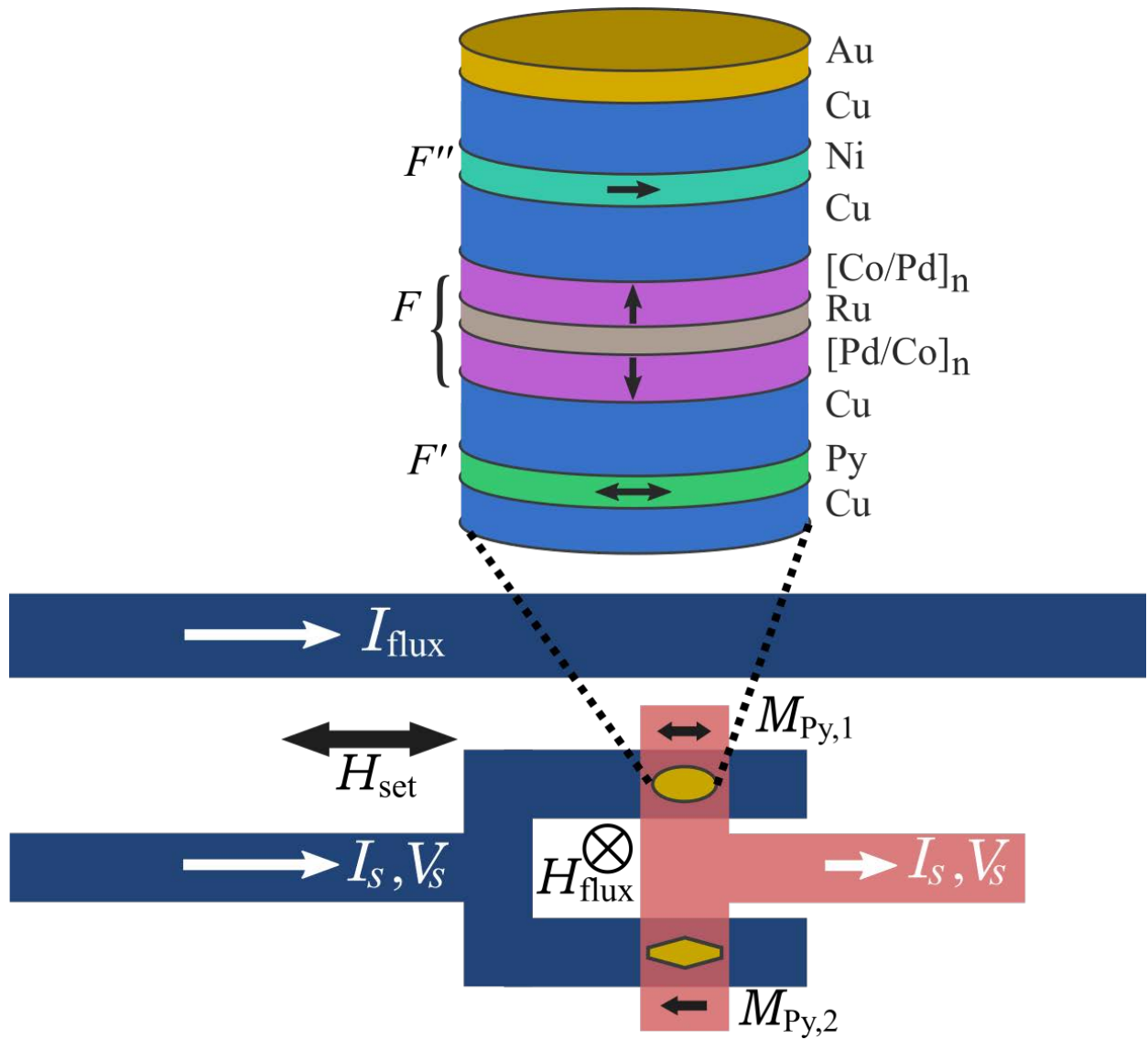

**Fig. 2. (A)** 3D plot of a minor loop for SQUID 2A-4: critical current vs $I_{flux}$ and $H_{set}$. The critical current plotted, $I_{c,Avg}$, is the average of the critical currents in the positive and negative current directions, $I_{c,Avg}$ = ($I_{c+}$ + $|I_{c-}|$)/2 (see panel B). Before any measurements are made, $\mu_0 H_{set}$ is set to -150 mT to initialize the magnetizations of the Ni and Py layers in both junctions. With $H_{set}$ =0, $I_c$ is measured as a function of $I_{flux}$ and exhibits oscillations with a period of about 1.5 mA, corresponding to the flux quantum, $\Phi_0$ = h/2e. Then $H_{set}$ is stepped in the positive direction (labeled "Up sweep" in the figure), returning to zero after each step for sample measurement. The SQUID oscillations exhibit a horizontal shift of ½$\Phi_0$ at $\mu_0 H_{set}$ = +2.4 mT, indicating that one of the Josephson junctions has changed its ground-state phase by $\pi$. The SQUID remains in that state as $H_{set}$ is increased further, but increasing $H_{set}$ too far causes the second junction to switch; the data shown here stop before that occurs. Next, $H_{set}$ is stepped in the negative direction (labeled "Down sweep") until $\mu_0 H_{set}$ = -2.8 mT, where the SQUID switches back to the original state. **(B)** Current-voltage characteristics obtained at $I_{flux}$ = -0.2 mA for the two magnetic states: in the $\pi$ state with maximum $I_c$ (green symbols) and in the initial 0 state with minimum $I_c$ (purple symbols). The solid green and purple lines are fits to the I-V curves with the Ivanchenko-Zil'bermann (IZ) function as described in the Supplementary Material, while the red and blue dashed lines are fits to the simpler square-root function used to obtain the data in panels (A) and (C). The latter fits give values of $I_c$ about 20% lower than the former, as shown by the $I_{c+}$ and $I_{c-}$ labels. **(C)** Repeated switching between the P and the AP-states at $I_{flux}$ = -0.2 mA. The histogram shows the measured values of $I_c^{avg}$ in the two states while the magnetic field was toggled between +2.8 mT and -3.2 mT one thousand times.

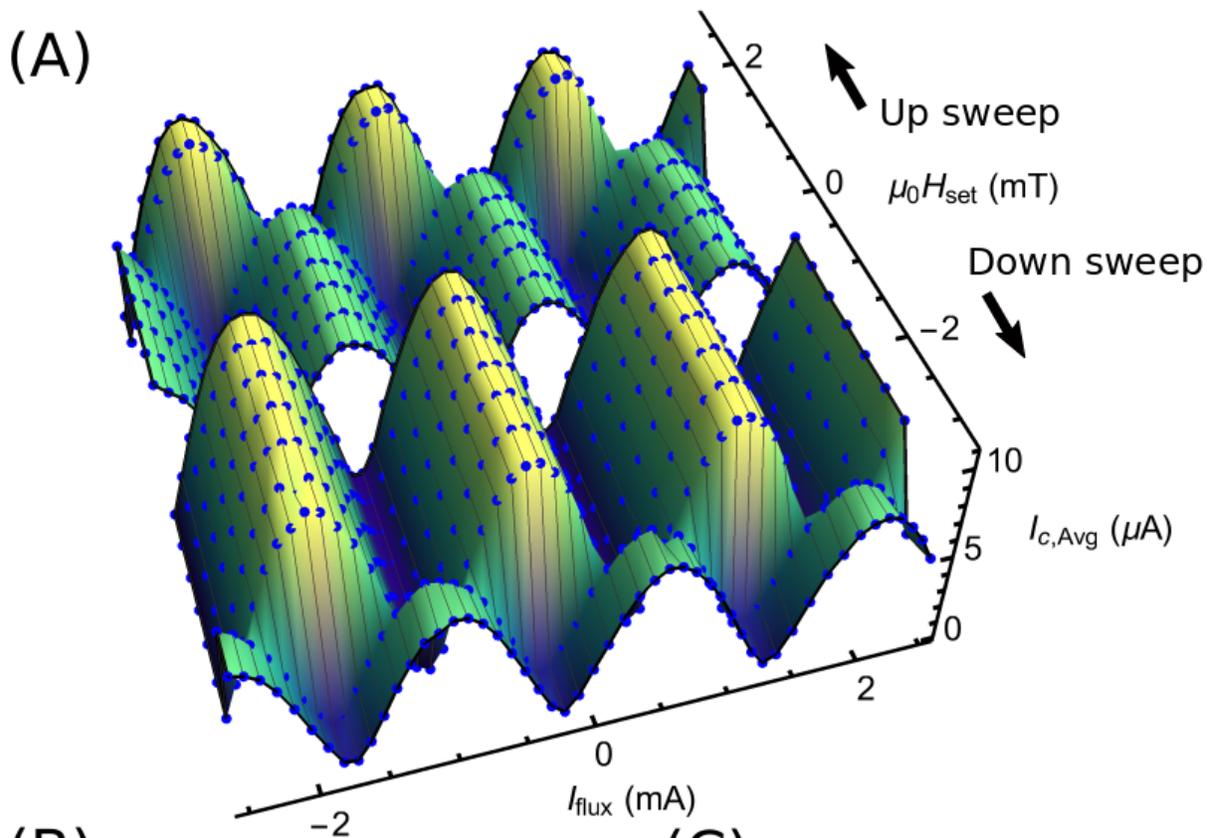

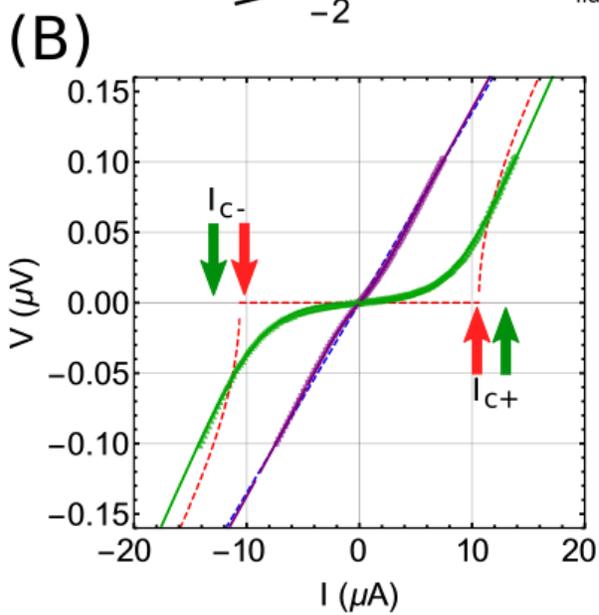

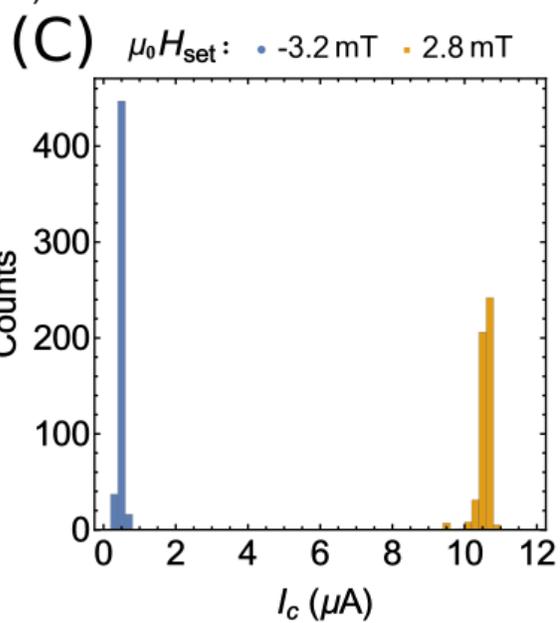

**Fig. 3.** Quantitative fits to the data for SQUID 2A-4; plot of $I_{c+}$ and $I_{c-}$ vs $I_{flux}$ for both magnetic states. To increase the accuracy of the data analysis, the values of $I_{c+}$ and $I_{c-}$ shown in this figure were obtained by fitting the Ivanchenko-Zil'berman function to the raw I-V curves (see Figure 2(B) and the Supplementary Material). The solid lines are least-squares fits to the data of standard SQUID theory [25]. Values of the SQUID loop inductance and critical currents of the two Josephson junctions obtained from the fits are given in the Supplementary Material.

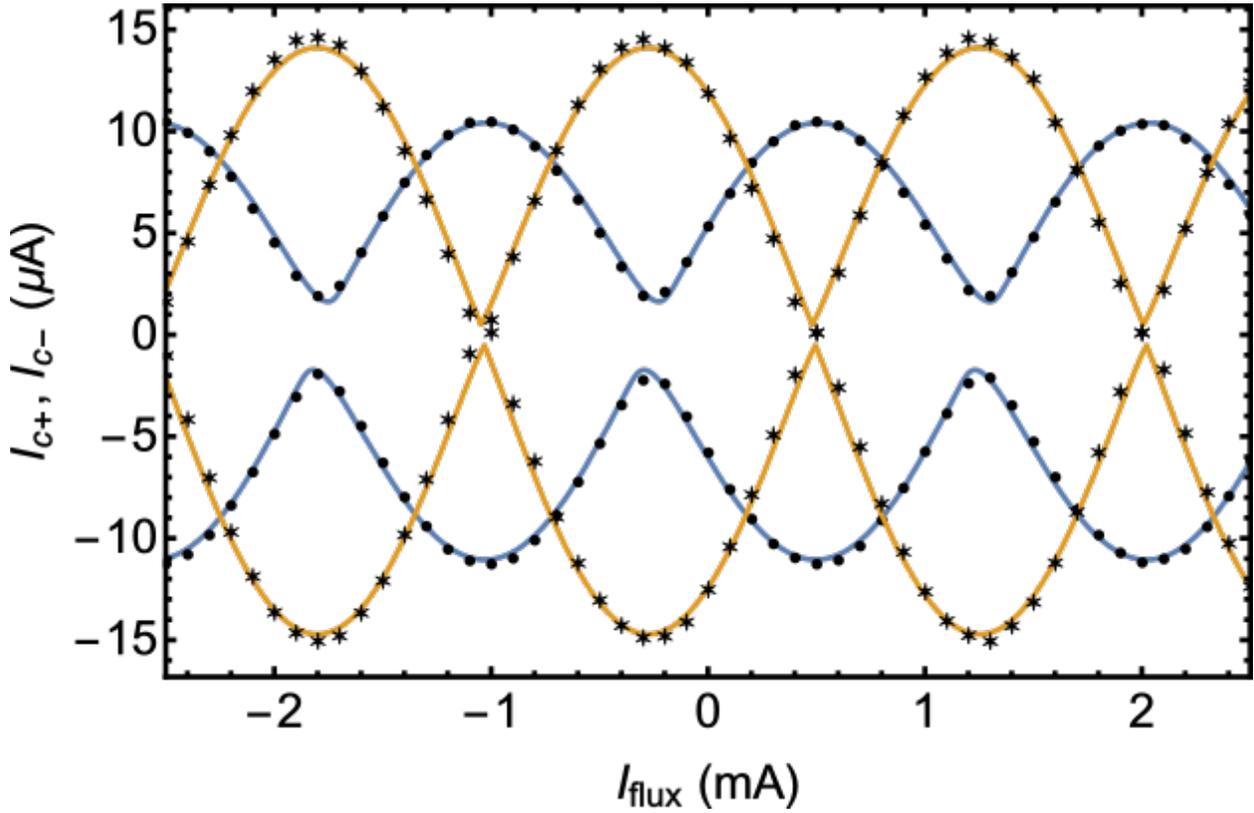

## Supplementary Materials

**Materials and Methods:**

The sample fabrication procedure was described in detail in [14]; here we summarize the main steps. The bottom leads consist of a multilayer of the form:
[Nb(25)/Al(2.4)]$_3$/Nb(20)/Au(2)/Cu(2)/Py(1.25)/Cu(4)/[Pd(0.9)/Co(0.3)]$_n$/Ru(0.95)/[Co(0.3)/Pd(0.9)]$_n$/Cu(4)/Ni(1.6)/Cu(7)/Au(2), where all thicknesses are in nm.  The [Nb/Al] multilayer base is used in place of pure Nb because it has a smoother surface and leads to better magnetic switching behavior of the free Py layer (see [14] and references therein).  The [Pd/Co] multilayers have strong perpendicular magnetic anisotropy (PMA), and the Ru(0.95) spacer couples the two PMA multilayers into a synthetic antiferromagnet (SAF).  The Cu(4) spacers decouple adjacent magnetic layers, while the bottom Cu(2) spacer facilitates growth of fcc Py on top of bcc Nb.  The multilayer is sputtered in a 7-gun high-vacuum sputtering system with a base pressure below $2\times10^{-8}$ Torr, with the substrate maintained at a temperature between -30°C and -15°C.  Because the multilayer contains 9 different materials while the system contains only 7 guns, the sputtering was interrupted after the first Au(2) layer and the system was opened to replace the Nb and Al guns with the Co and Pd guns.  During the gun exchange the samples were enclosed in a plastic bag filled with flowing $N_2$ gas to limit contamination.  After the system was closed, it was pumped down overnight, and the Au(2) was ion milled away prior to deposition of the remainder of the stack.  The entire bottom multilayer described above was deposited through a photolithographic stencil (with S1813 photoresist) to define the dimensions of the bottom leads using the lift-off process.

Following lift-off, e-beam lithography and Ar ion milling were used to define the junction areas, using the negative e-beam resist ma-N2401 as the ion mill protective mask.  Immediately following ion milling, 50 nm of SiO$_x$ were deposited by thermal evaporation for electrical isolation.  Then the samples were ion milled at glancing angle from two directions to break through the SiO$_x$ sidewalls around the junctions.  Subsequent lift-off of the ma-N2401 was performed in warm PG Remover with the aid of gentle wiping with a cotton swab.  Following lift-off, the sample was subjected to an $O_2$ plasma "descum" process to ensure complete removal of the ma-N2401 from the tops of the junctions.  Finally, the top lead pattern was defined with another photolithography step.  The protective Au(2) layer was ion milled away immediately prior to sputtering the top Nb(150)/Au(10) electrode.  Sample fabrication was completed by lift-off of the photoresist.

For measurement, NbTi wires are attached to the Nb/Au pads using pressed indium.  The samples are immersed in liquid He.  Current is provided by a battery-powered low-noise current source, while the voltage is measured using a SQUID-based self-balancing potentiometer circuit with a voltage noise of a few pV/$\sqrt{}$Hz [26].  Current-voltage (I-V) characteristics are typical of overdamped Josephson junctions [27], but the I-V curves exhibit substantial rounding for currents less than the critical current, $I_c$, due to environmental and instrumental noise, as shown in Fig. 2(B).  Rounded I-V curves are well described by the Ivanchenko-Zil'berman (IZ) function [28,29,14], while less-precise estimates of $I_c$ can be obtained by fitting the data to the standard form for overdamped Josephson junctions: $V = R\times\text{Real}\{(I^2-I_c^2)^{1/2}\}$. Because fitting I-V curves with the IZ function is time-consuming, we carry out the initial data analysis using the simpler square-root fits, then use the IZ function to fit the data sets that will be subjected to further quantitative analysis.

One can extract quantitative information from the oscillatory $I_c(I_{flux})$ data by fitting them with the standard SQUID theory [25]. In principle, such fits provide estimates of the four SQUID parameters – the critical currents of the two Josephson junctions, $I_{c1}$ and $I_{c2}$, and the inductances of the two arms of the SQUID, $L_1$ and $L_2$ – and an overall phase shift, $\phi$. Our SQUIDs are in the low-inductance limit where $(I_{c1} + I_{c2})(L_1 + L_2) \ll \Phi_0$ (or $\beta_L \ll 1$); in that limit, it is difficult to extract accurate inductance estimates from the fits. Nevertheless, we have good estimates of the inductances from numerical simulations of the SQUID structure using the FastHenry software, as well as from previous work with similar SQUIDs containing Josephson junctions with much larger critical currents [17]. Since fluctuations in the fitted values of $L_1$ and $L_2$ tend to correlate with fluctuations in the fitted values of $I_{c1}$ and $I_{c2}$, we obtained our most consistent fitting results by fixing the values of the two inductances to $L_1 = L_2 = 4.5$ pH, and letting only $I_{c1}$, $I_{c2}$, and $\phi$ vary in the fitting procedure. The results of fitting all 7 data sets, in both magnetic states, are shown in Table I. The most important results are in the last column, which shows the change in $\phi$ between the two magnetic states for each of the 7 SQUIDs, in units of $2\pi$. All phase changes $\Delta\phi$ are very close to 0.5, corresponding to a phase shift of $\pi$ for one of the two junctions in the SQUID. Small deviations from $\pi$ are probably due to changes in the SQUID flux coming from the switching Py layer [17]. Given the robust results for $\Delta\phi$, we are confident that only one of the two Josephson junctions in the SQUID changes its magnetic state at the transitions visible in Fig. 2. Table 1 confirms the general picture that $I_{c1}$ changes much more than $I_{c2}$ in all samples. We believe that the changes in $I_{c2}$ shown in the table indicate limitations of the fitting procedure for SQUIDs with such small critical currents.

As discussed above, the small critical currents of these samples lead to significant rounding of the I-V curves. That rounding is caused by fluctuations in the electromagnetic environment of the junctions, due either to temperature, the measurement apparatus, or interference from sources external to the cryostat. Such fluctuations not only cause rounding, but they may also decrease the amplitude of the measured critical current. While the qualitative conclusions of this work are immune to such considerations, the quantitative accuracy of the analysis is not. To ascertain the extent to which external interference may have influenced the maximum measured critical currents of our samples, we measured one sample in a variable-temperature cryostat with heavily filtered electrical lines. That cryostat is not equipped with the SQUID-based self-balancing potentiometer circuit, so we measured dV/dI vs I of the junctions using an ac technique with a lock-in amplifier. The dV/dI vs I data were then integrated to obtain the V vs I curves shown in Fig. S1(A), acquired at T = 2.0 K for several different values of $I_{flux}$. Those I-V curves were fit with the square-root function (dashed lines), and a plot of $I_c$ vs $I_{flux}$ is shown in Fig. S1(B). Fig. S1(C) shows a plot of $I_c$ vs temperature over the range 1.2 – 5.6 K. The values of $I_c$ shown in this figure, as well as the extent of rounding of the I-V curves, are comparable to those obtained using the SQUID-based circuit. We conclude that the values of $I_c$ measured in this work are not strongly attenuated by external interference. (Attempts to fit the data in Fig. S1(A) with the Ivanchenko-Zil'berman function were unsuccessful due to small slope offsets in the data.)

**Figure S1:** Critical current versus flux-line current and temperature for SQUID 2A-1 measured in a variable-temperature cryostat with heavily-filtered lines. **(A)** Data of dV/dI vs I were acquired at T = 2.0K using a lock-in amplifier technique, and were converted into I-V curves by numerical integration. The I-V curves display similar thermal rounding as the curves obtained using the SQUID-based self-balancing potentiometer circuit at 4.2 K, shown in Figure 2(B). From top to bottom, each I-V curve corresponds to a different value of the applied flux current, starting at -1.0 mA and increasing by 0.2 mA for each successive curve. The curves are successively offset vertically by 0.25 µV for clarity. The dashed lines represent fits to the simple square-root function. (Fitting to the more accurate Ivanchenko-Zil'berman function was unsuccessful for a few of these curves due to the nonzero slope at zero applied current.) **(B)** Plot of critical current at 2.0 K versus $I_{flux}$. The dashed line is a guide the eye. **(C)** Measurements of critical current versus temperature, $I_c(T)$, with $I_{flux}$ =0. While $I_c$ increases with decreasing T, it is not substantially larger at 1-2 K than the previous measurements at 4.2 K. Temperature values above 4.2 K have large uncertainty due to extrapolation of the thermometer calibration curve.

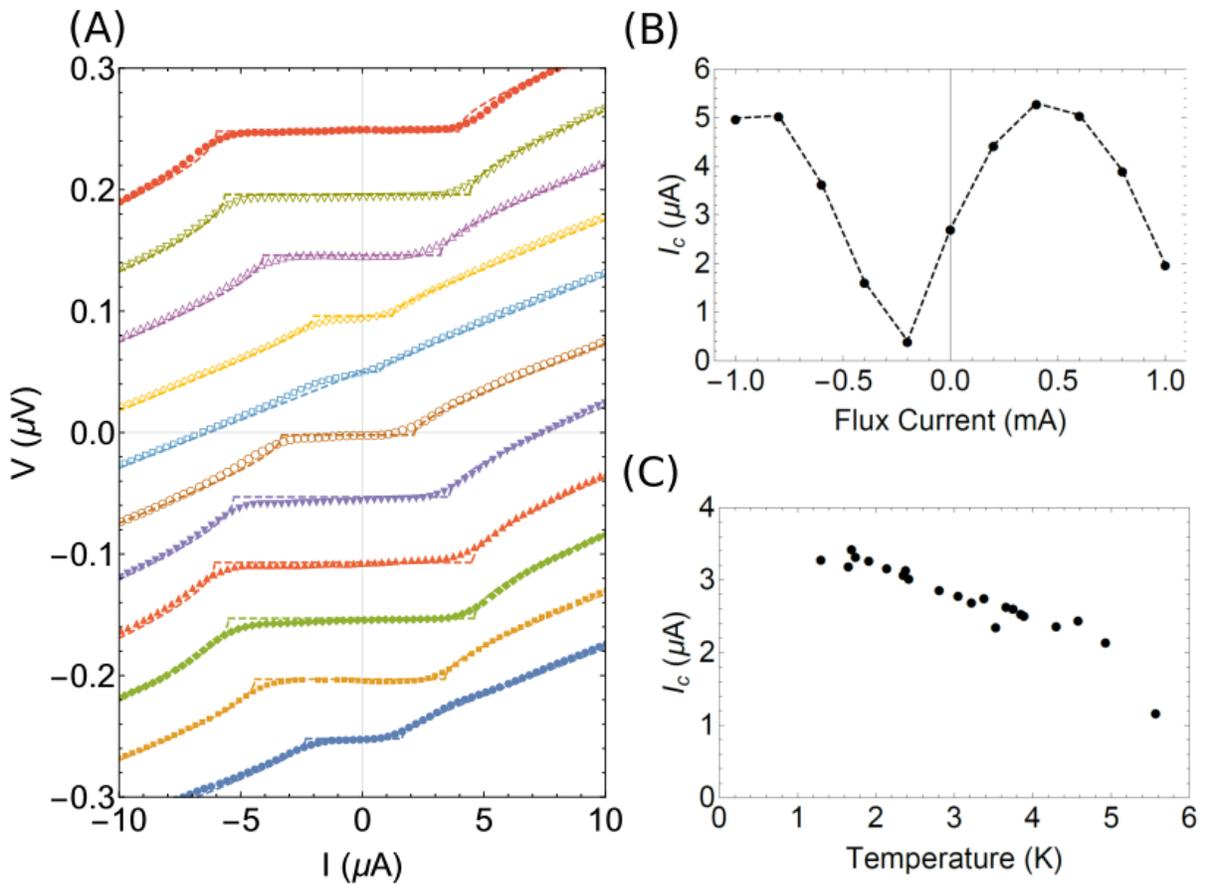

**Table 1:** Summary of the seven spin-triplet SQUID samples measured. The first two characters in the sample name, e.g. "2A", refer to the chip, while the final number refers to the specific device on the chip. (Each chip contains four SQUIDs.) The SQUID oscillation curves were fit to standard SQUID theory while keeping the total inductance fixed to the nominal value of 9 pH. The last column shows that the phase differences between the two magnetic states in all the SQUIDs are very close to $\pi$. (The shift is given in units of $2\pi$.) Moreover, from the fit parameters we can extract approximate values of the critical currents in the two junctions. Typically, we find that between the two states only the critical current in the elliptical junction ($I_{c1}$) changes in magnitude, while the critical current in the hexagonal junction ($I_{c2}$) is approximately constant. The value $2n$ is the total number of [Pd/Co] bilayers in the central F layer. (Note that the uncertainties in the values of $I_{c1}$ an $I_{c2}$ derived from the fits are too small, as the value of $I_{c2}$ for the non-switching junction appears to change between the two magnetic states. We believe that this is a generic feature of fits to SQUID data for SQUIDs in the low-inductance limit, $\beta_L \ll 1$.) The data on which these fits were obtained can be found in the Appendix of [30].

| SQUID name | 2n | State | $I_{c1}$ (µA) | $I_{c2}$ (µA) | $\Delta\phi/2\pi$ |
|---|---|---|---|---|---|
| 2A-1 | 4 | 1 | 6.65±0.08 | 4.20±0.08 | 0.491±0.005 |
|  |  | 2 | 6.90±0.12 | 4.18±0.12 |  |
| 2A-2 | 4 | 1 | 5.66±0.10 | 5.61±0.10 | 0.542±0.004 |
|  |  | 2 | 4.09±0.14 | 4.02±0.14 |  |
| 2A-3 | 4 | 1 | 4.53±0.12 | 4.56±0.12 | 0.480±0.004 |
|  |  | 2 | 6.88±0.07 | 4.67±0.07 |  |
| 2A-4 | 4 | 1 | 4.56±0.04 | 6.16±0.04 | 0.509±0.003 |
|  |  | 2 | 7.30±0.15 | 7.08±0.15 |  |
| 3A-3 | 6 | 1 | 0.80±0.30 | 0.79±0.30 | 0.519±0.010 |
|  |  | 2 | 1.99±0.02 | 1.34±0.02 |  |
| 4A-1 | 6 | 1 | 0.60±0.33 | 3.16±0.33 | 0.618±0.005 |
|  |  | 2 | 1.45±0.03 | 3.87±0.03 |  |
| 4A-2 | 6 | 1 | 1.32±0.04 | 2.41±0.04 | 0.493±0.004 |
|  |  | 2 | 3.01±0.01 | 1.61±0.01 |  |